\documentclass[onecolumn,prb,notitlepage,floats,superscriptaddress,amsmath,amssymb]{revtex4-1}

\usepackage[version=3]{mhchem} 
\usepackage{bm}
\usepackage[utf8]{inputenc}
\usepackage[T1]{fontenc}
\usepackage{graphicx}
\usepackage{upgreek}
\usepackage{color}
\usepackage{ulem}
\usepackage{hyperref} 
\hypersetup{colorlinks,citecolor=blue, filecolor=blue ,linkcolor=blue , urlcolor=blue, pdftex}
\usepackage{sidecap}
\usepackage{amssymb}

\def \FUW{Institute of Experimental Physics, Faculty of Physics, University of Warsaw, ul. Pasteura 5, 02-093 Warsaw, Poland}
\def \Wroclaw{Department of Semiconductor Materials Engineering, Wrocław University of Science and Technology, ul. Wybrzeże Wyspiańskiego 27, 50-370 Wrocław, Poland}

\begin{document}
	
	\title{Anisotropic Optical And Vibrational Properties Of GeS}

\author{Natalia Zawadzka}
\affiliation{\FUW}
\author{\L{}ucja Kipczak}
\affiliation{\FUW}
\author{Tomasz Wo\'zniak}
\affiliation{\Wroclaw}
\author{Katarzyna Olkowska-Pucko}
\affiliation{\FUW}
\author{Magdalena Grzeszczyk}
\affiliation{\FUW}
\author{Adam Babi\'nski}
\affiliation{\FUW}
\author{Maciej R. Molas}
\email{maciej.molas@fuw.edu.pl}
\affiliation{\FUW}

\begin{abstract}
	
The optical response of bulk germanium sulfide (GeS) is investigated systematically using different polarization-resolved experimental techniques, such as photoluminescence (PL), reflectance contrast (RC), and Raman scattering (RS). It is shown that while the low-temperature ($T$=5 K) optical band-gap absorption is governed by a single resonance related to the neutral exciton, the corresponding emission is dominated by the disorder/impurity- and/or phonon-assisted recombination processes. Both the RC and PL spectra are found to be linearly polarized along the armchair direction. The low and room ($T$=300 K) temperature RS spectra consist of six Raman peaks identified with the help of Density Fuctional Theory (DFT) calculations: A$^1_{\textrm{g}}$, A$^2_{\textrm{g}}$, A$^3_{\textrm{g}}$, A$^4_{\textrm{g}}$, B$^1_{\textrm{1g}}$, and B$^2_{\textrm{1g}}$, which polarization properties are studied under four different excitation energies. We found that the polarization orientations of the A$^2_{\textrm{g}}$ and A$^4_{\textrm{g}}$ modes under specific excitation energy can be useful tools to determine the GeS crystallographic directions: armchair and zigzag.
	
\end{abstract}

\maketitle

\section{Introduction}
Two-dimensional (2D) layered van der Waals (vdW) semiconductors, such as transition metal dichalcogenides ($e.g.$ MoS$_2$, WSe$_2$), have appeared as a fascinating class of materials for exploring novel excitonic phenomena.~\cite{mak2010, AroraWSe2, AroraMoSe2, MolasWS2} Inspired by these exciting achievements, a new group of emerging vdW  semiconductors, $i.e.$ group-IV monochalcogenides MX (where M = Ge, Sn or Pb and X = S, Se or Te), has attracted increasing attention due to their anisotropic optical properties. They originate from a low-symmetry orthorhombic crystal structure, analogous to black phosphorous (BP), which is widely investigated.~\cite{Ling2016, ribeiro2015, molas2021} Moreover, the family of MX materials exhibits high carrier mobility, larger for monolayers as compared to bulk,~\cite{Li2016} which can lead to  potential applications in angle-resolved opto-electronics.

In this work, we investigate anisotropic optical and vibrational properties of GeS with the aid of the polarization-resolved photoluminescence (PL), reflectance contrast (RC), and Raman scattering (RS. The low-temperature ($T$=5 K) optical band-gap absorption is governed by a single resonance related to the neutral exciton, the corresponding emission is dominated by disorder/impurity- and/or phonon-assisted recombination processes. Both the RC and PL spectra are found to be linearly polarized along the armchair direction of the crystal. The low and room ($T$=300 K) temperature RS spectra consist of six Raman peaks, identified with the help of Density Fuctional Theory (DFT) calculations. Their polarization properties are studied under four different excitation energies. We found that the polarization orientations of the A$^2_{\textrm{g}}$ and A$^4_{\textrm{g}}$ modes under specific excitation energy can be used to distinguish between armchair and zigzag crystallographic directions in GeS crystals.

\section{Materials and Methods}
\subsection{Samples}
A bulk-like flake of GeS was placed on a Si/(90 nm) SiO$_2$ substrate by polydimethylsiloxane (PDMS)-based exfoliation of bulk crystals purchased from HQ Graphene. The flake of interest was initially identified by visual inspection under an optical microscope then subjected to atomic force microscopy.

\subsection{Experimental techniques}
The PL spectra measured under laser excitation of $\lambda$= 660 nm (1.88 eV). The RS measurements were performed using illumination with a series of lasers: $\lambda$= 488 nm (2.54 eV), $\lambda$= 515 nm (2.41 eV), $\lambda$= 561 nm (2.21 eV), $\lambda$= 633 nm (1.96 eV). The excitation light in those experiments was focused by means of a 50x long-working distance objective with a 0.55 numerical aperture (NA) producing a spot of about 1 $\mu$m diameter. The signal was collected via the same microscope objective (the backscattering geometry), sent through a 0.75 m monochromator, and then detected by using a liquid nitrogen cooled charge-coupled device (CCD) camera. To detect low-energy RS below 100 cm$^{-1}$ from the laser line, a set of Bragg filters was implemented in both excitation and detection paths. In the case of the RC studies, the only difference in the experimental setup with respect to the one used for recording the PL and RS signals concerned the excitation source, which was replaced by a tungsten halogen lamp. The light from the lamp was coupled to a multimode fiber of a 50 $\mu$m core diameter, and then collimated and focused on the sample to a spot of about 4 $\mu$m diameter. 
All measurements were performed with the samples placed on a cold finger in a continuous flow cryostat mounted on x–y manual positioners. The excitation power focused on the sample was kept at 50 $\mu$W during all measurements to avoid local heating.

The polarization-resolved PL and RC spectra were analyzed by the motorized half-wave plate and a fixed linear polarizer mounted in the detection path. In contrast, the polarization-sensitive RS measurements were performed in two co- (XX) and cross-linear (XY) configurations, which correspond to the parallel and perpendicular orientation of the excitation and detection polarization axes, respectively. The analysis of the RS signal was done using a motorized half-wave plate, mounted on top of the microscope objective, which provides simultaneous rotation of polarization axis in the XX and XY configurations.

\subsection{Theoretical calculations}
DFT calculations were performed in Vienna Ab initio Simulation Package \cite{Kresse1996} with Projector Augmented Wave method \cite{Kresse1999} and Perdew-Burke-Ernzerhof parametrization \cite{Perdew1996} of general gradients approximation of an exchange-correlation functional. A plane wave basis energy cutoff of 550 eV and a $\Gamma$-centered Monkhorst-Pack k-grid of 12$\times$10$\times$4 were found sufficient to converge the lattice constants up to 0.001 \AA. Geometrical parameters were optimized until the interatomic forces and stress tensor components were lower than 10$^{-5}$ eV/\AA~and 0.01 kbar, respectively. The interlayer vdW interactions were taken into account by the use of Grimme's D3 correction \cite{Grimme2010}. Phonon dispersion of GeS was calculated within Parliński-Li-Kawazoe method \cite{Parlinski1997}, as implemented in Phonopy software \cite{Togo2015}. The 4$\times$4$\times$2 supercells were employed to find the interatomic force constants within the harmonic approximation. The irreducible representations of Raman active phonon modes at $\Gamma$ point were determined with the use of spglib library \cite{Togo2018}.

\section{Results}
\subsection{Crystallographic structure}

\begin{figure}[b!]
\centering
\includegraphics[width=0.549\linewidth]{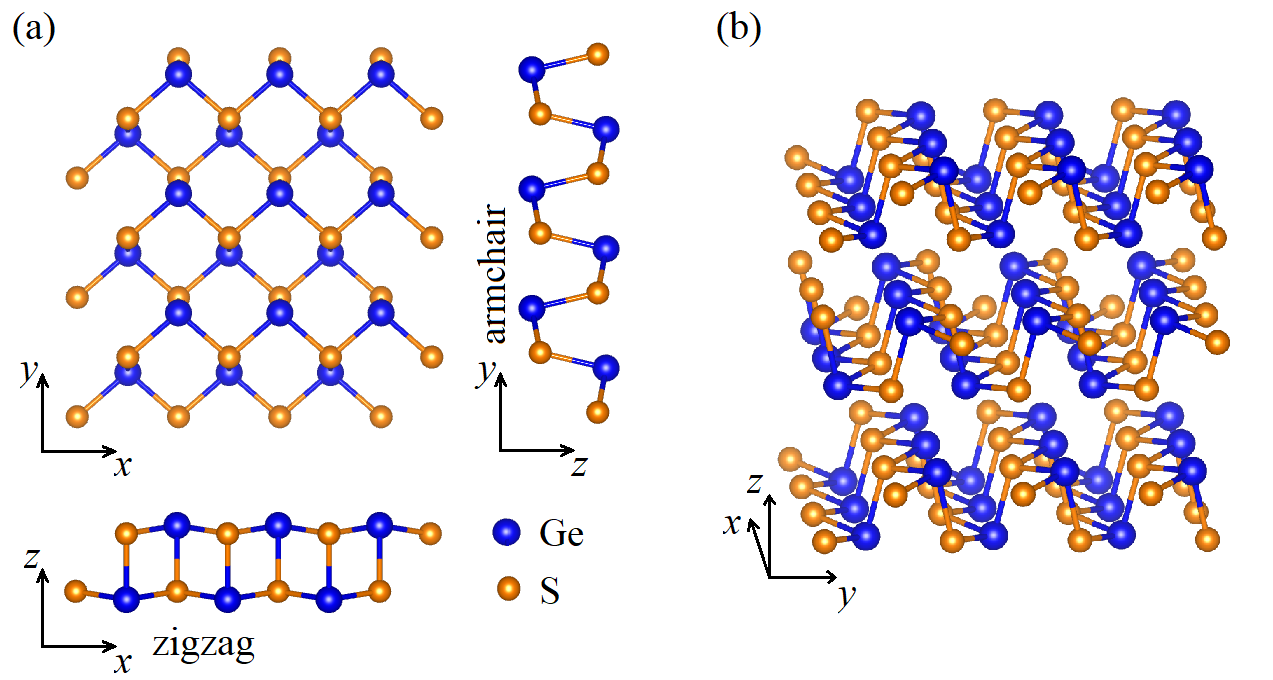}
\caption{(a) Top and side views of the GeS crystal structure for a single layer. The armchair and zigzag directions are shown in relation to the crystal orientation and lattice parameters. (b) Geometrical structure of multilayer GeS. \label{fig1}}
\end{figure}

GeS is a layered material, which crystallizes in a distorted orthorhombic structure (D$^16_{2h}$), as shown in  Figure~\ref{fig1}. That form comprising eight atoms in the primitive unit cell has been proven to be dynamically and thermally stable at room temperature~\cite{Siemers}. The puckered honeycomb lattice of GeS has an anisotropic crystal structure characterized by the two orthogonal armchair ($x$) and zigzag ($y$) directions, denoted in Figure~\ref{fig1}(a). The stack of consecutive layers in the perpendicular direction ($z$) in respect to the $xy$ plane is presented in Figure~\ref{fig1}(b). Note that the crystallographic structure of GeS is analogous to the BP one.\cite{molas2021}

\subsection{Optical properties}

\begin{figure}[t]
			\centering
			\includegraphics[width=0.5\linewidth]{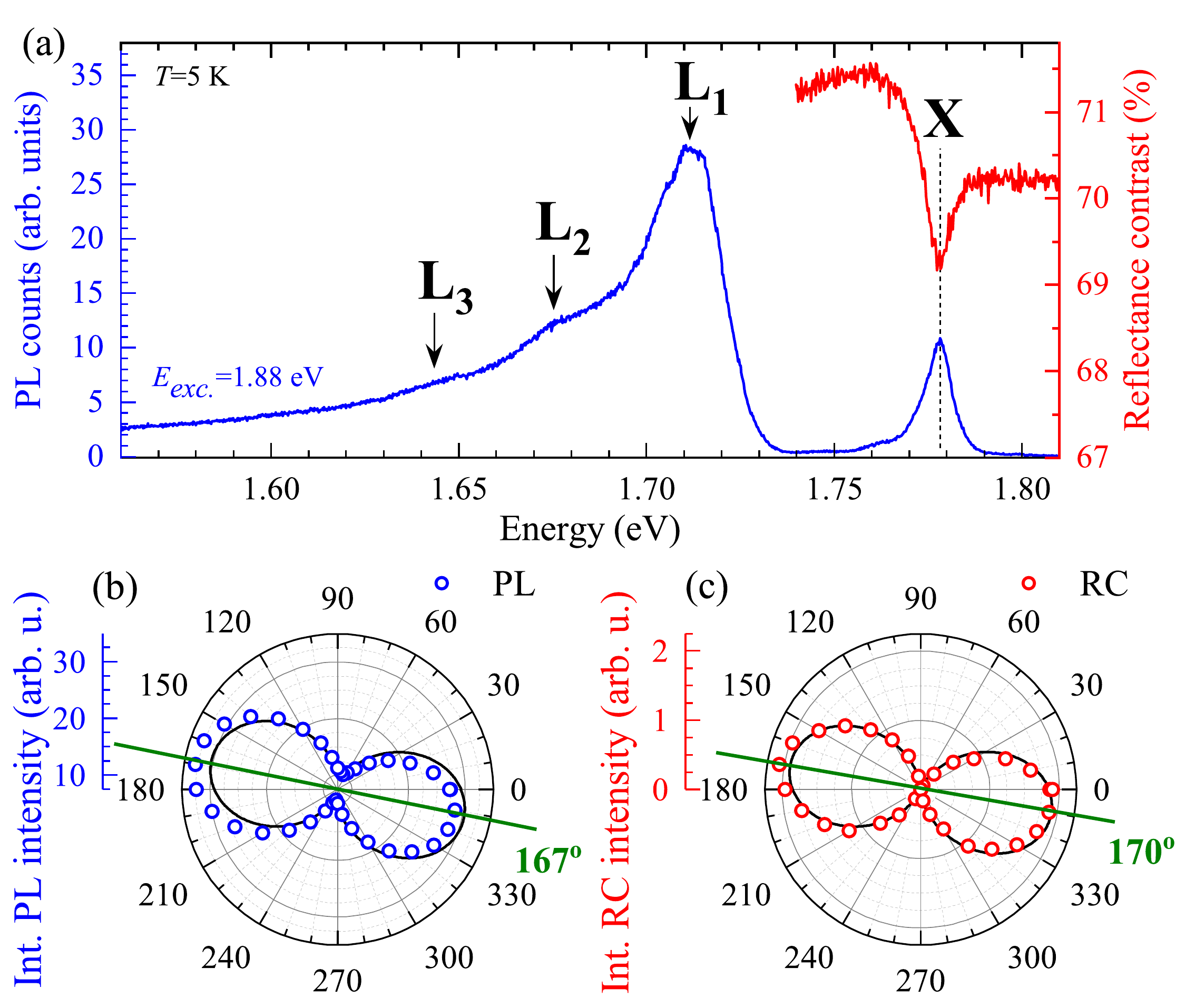}
		\caption{(a)The low-temperature ($T$=5 K) PL (blue curve) and RC (red curve) spectra measured on the GeS flake. Polar plots of the integrated intensities of the X transitions from the (b) PL and (c) RC spectra. \label{fig2}}
\end{figure}

 \begin{figure}[b!]
\centering
\includegraphics[width=0.61\linewidth]{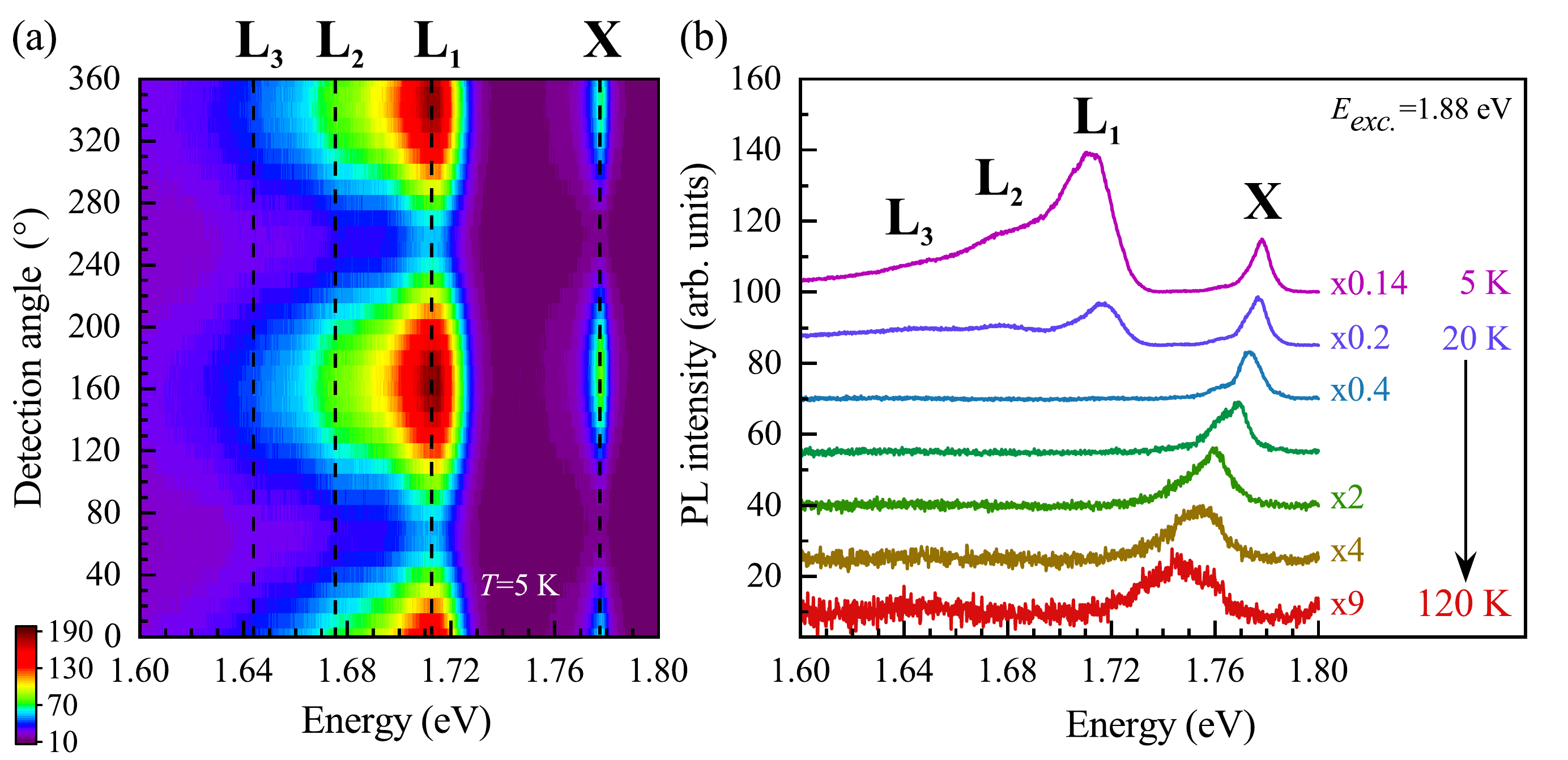}
\caption{(a) False-color map of low-temperature PL spectra of GeS as a function of the detection angle of linear polarization under 1.88 eV laser light excitation. (b) Temperature evolution of the PL spectra measured on GeS. The spectra are vertically shifted for clarity and some of them are multiplied by scaling factors for more clarity.  \label{fig3}}
\end{figure}   

Figure~\ref{fig2}(a) presents the low-temperature ($T$=5~K) PL and RC spectra. 
The PL spectrum consists of four emission lines, denoted as X, L$_1$, L$_2$ and L$_3$. 
In contrast, the corresponding RC spectrum consists of a single resonance, which energy coincides with the X emission line. To examine the origin of the X transition, we carried out the polarization resolved measurements of PL and RC spectra. 
The extracted polarization dependencies of the X transitions are presented in Figures~\ref{fig2}(b) and (c). Solid lines in the Figs represent fits of the evolution of the PL/RC intensity as a function of light polarization. A dichroic relation of polarized light using Malus law was used:  $I(\theta)$:\cite{Hecht}
\begin{equation}
I(\theta)=Acos^2 (\theta-\phi),
\end{equation}
where $A$ is the amplitude of the emission/absorption transition and $\phi$ represents the phase of polarization dependence. It is seen that both the emission and absorption signals of the X transitions are linearly polarized along the same direction (167$^o$ and 170$^o$, respectively). The result confirms directly the same origin of the X feature apparent in both the PL and RC spectra, as previously reported independently for the emission\cite{Ho2016}, photoreflectance \cite{Oliva2020} or absorption\cite{tolloczko2020}. Moreover, according to Ref.~\citenum{tolloczko2020}, the X resonance can be related to the direct transition at the $\Gamma$ point of the Brillouin zone (BZ), which is linearly polarized along the armchair crystallographic direction.

In order to study the origin of L lines, the polarization-resolved and temperature evolution of the PL spectra was measured, see Figure~\ref{fig3}(a) and (b). 
As can be appreciated in the Figure~\ref{fig3}(a), the L$_1$, L$_2$ and L$_3$ lines are linearly polarized along the same armchair direction as the neutral exciton transition. Moreover, with increasing temperature, the low energy L$_1$, L$_2$ and L$_3$ peaks quickly disappear from the spectra. At $T$=40 K, only the neutral exciton contributes to the PL spectrum. The further increase of temperature leads to the typical redshift and the linewidth broadening of the neutral exciton, which can be observed up to 120~K. The observed temperature dependence of the L lines is very similar to the previously reported behavior of so-called "localized" excitons in monolayers of WS$_2$ and WSe$_2$ exfoliated on Si/SiO$_2$ substrates.\cite{Plechinger2015,Shang2015,AroraWSe2,Plechinger2016,Klopotowski2016,MolasWS2} Consequently, we can ascribe tentatively the L$_1$, L$_2$ and L$_3$ peaks to disorder/impurity- and/or phonon-assisted recombination processes.

\subsection{Vibrational properties}

\begin{figure}[b!]
\centering
\includegraphics[width=0.61\linewidth]{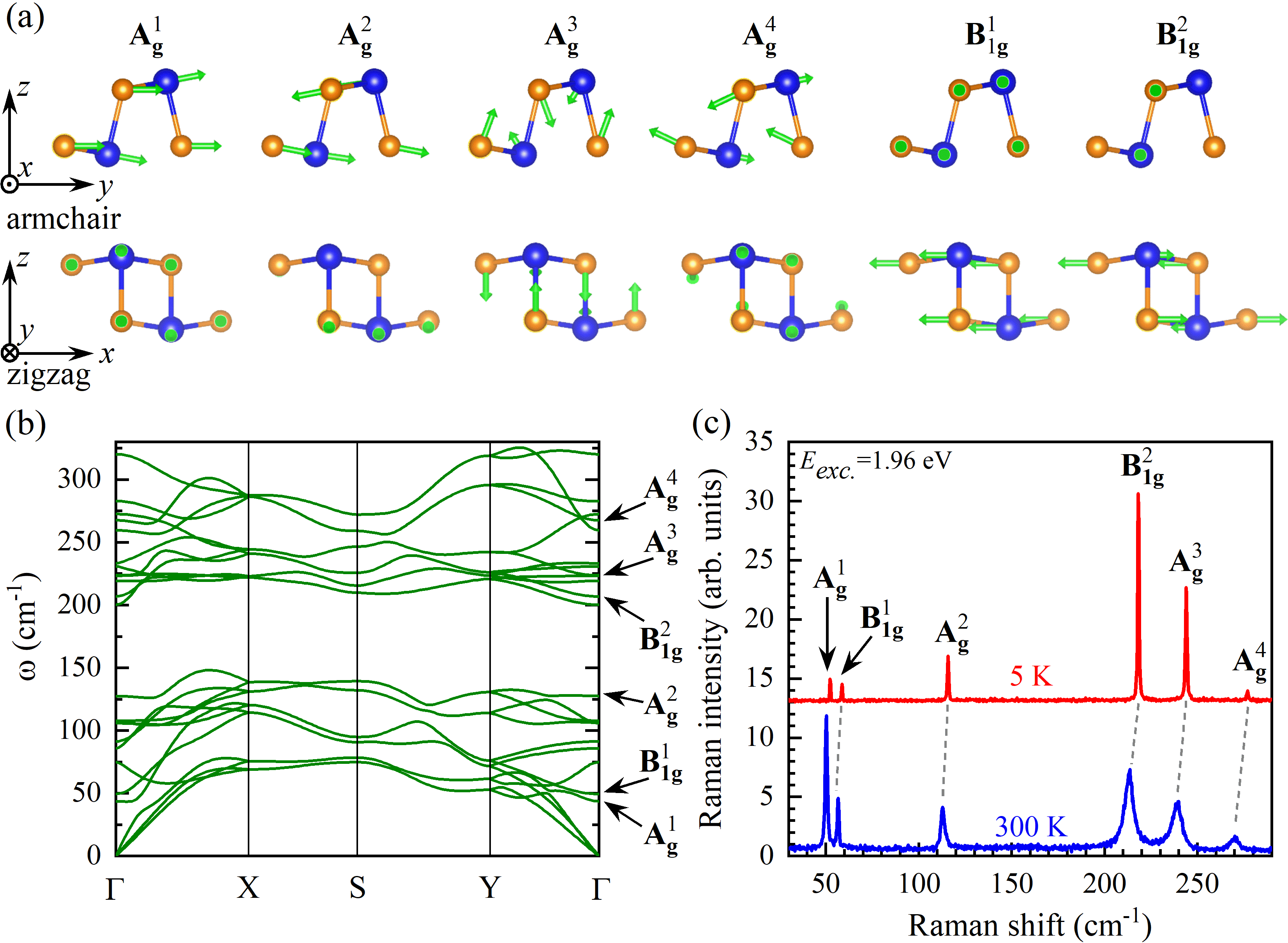}
\caption{(a) Atom displacements (green arrows) for the Raman-active modes. Axes indicate two view perspectives: armchair, and zigzag directions (b) The calculated phonon dispersion of GeS. (c) RS spectra measured on GeS at low ($T$=5 K) and room ($T$=300 K) temperatures. The spectra are vertically shifted for clarity. \label{fig4}}
\end{figure}

As GeS belongs to the point group D$_{2h}$, there are 12 Raman-active modes: 4A$_{g}$, 2B$_{1g}$, 2B$_{2g}$, 4B$_{3g}$. According to the polarization selection rules for the space group Pnma (No. 62) of GeS, four A$_{g}$ and two B$_{1g}$ phonon modes should be observed in backscattering geometry along the $z$ crystallographic direction. 
Their corresponding atomic displacements, denoted by green arrows, are presented in Figure~\ref{fig4}(a). 
The modes are classified according to their irreducible representations in the point group D$_{2h}$, and additionally numbered due to their increased Raman shift (top index). As can be seen in the Figure~\ref{fig4}(a), the A$_{g}$ modes presents atom movement mostly in the plane defined by the armchair and out-of-plane directions, while the B$_{1g}$ vibrations take place along zigzag direction. 
Figure~\ref{fig4}(b) presents the calculated phonon dispersion with marked A$_{g}$ and B$_{1g}$ phonon modes active in our experimental conditions. To verify the theoretical calculations, we measured the RS spectra of GeS at low ($T$=5 K) and room ($T$=300 K) temperatures. 
Both the RS spectra consist of six Raman modes: A$^1_{\textrm{g}}$, A$^2_{\textrm{g}}$, A$^3_{\textrm{g}}$, A$^4_{\textrm{g}}$, B$^1_{\textrm{1g}}$ and B$^2_{\textrm{1g}}$, which energies are in good agreement with theoretical calculations shown in Figure~\ref{fig4}(b). 
One notes the characteristic effect of temperature on the observed peaks. It is seen that both the redshift and the broadening of Raman peaks scales with the increasing phonon energy, $e.g.$ the temperature-induced change of the energy and linewidth of the A$^1_{\textrm{g}}$ mode is much smaller as compared to the A$^3_{\textrm{g}}$.

\begin{figure}[t]	
	\centering
	\includegraphics[width=0.61\linewidth]{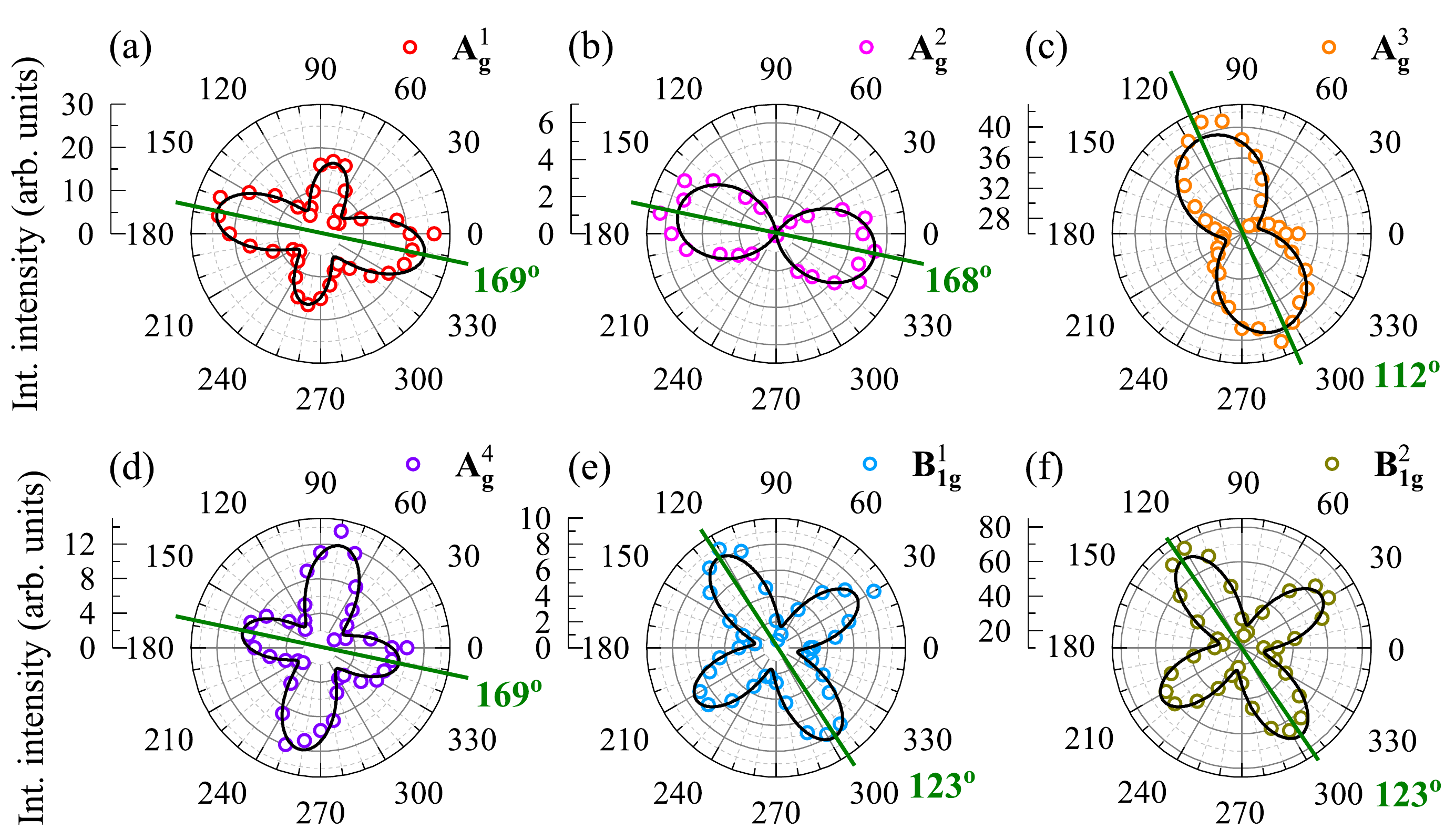}
	\caption{Polar plots of the integrated intensities of the phonon modes measured on GeS at $T$=300 K under 1.96 eV excitation. The green lines on polar plots are along polarization axes of modes. \label{fig5}}
\end{figure}  
 
 \begin{figure}[b!]
 	\centering
 	\includegraphics[width=0.61\linewidth]{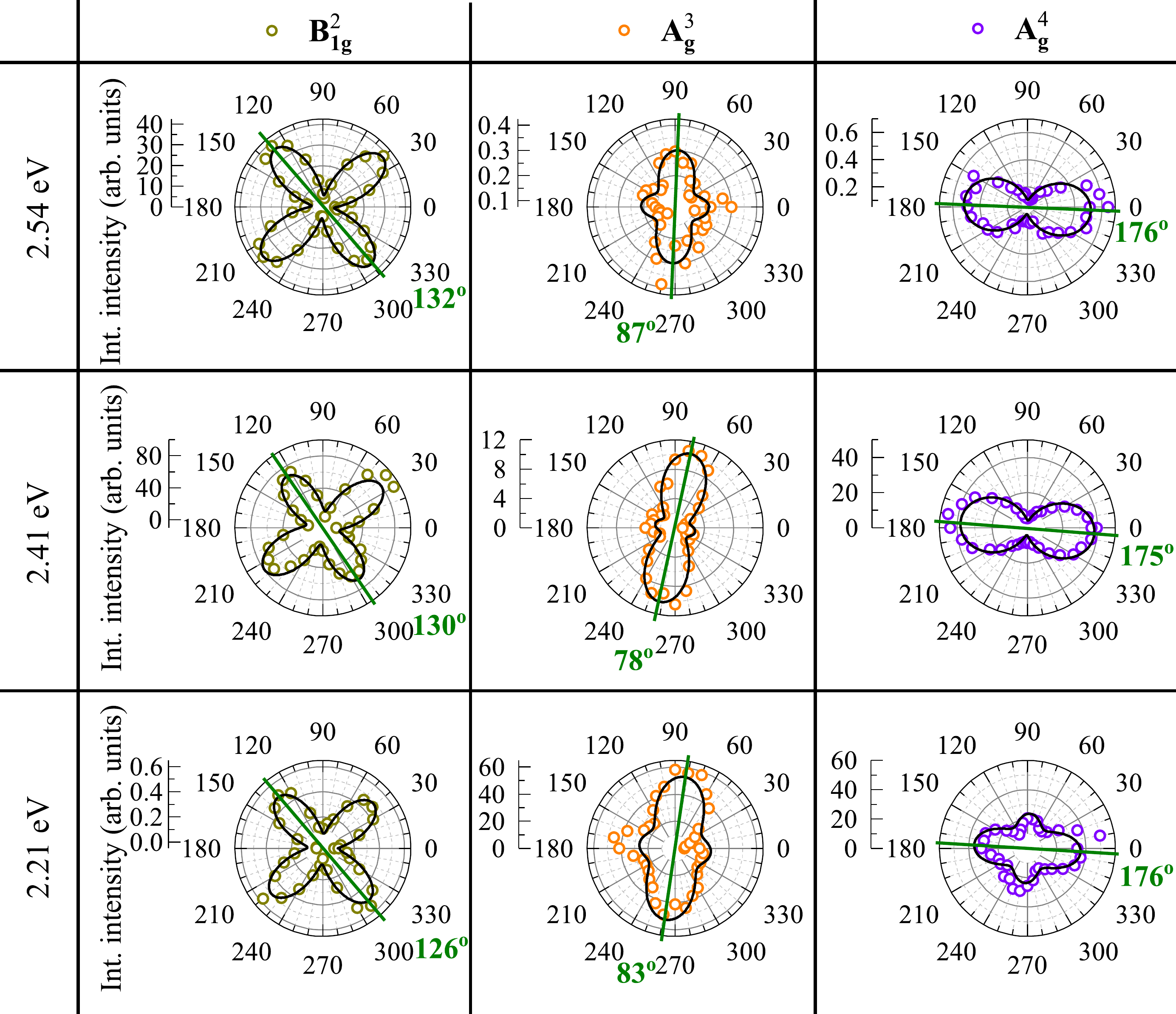}
 	\caption{Polar plots of the integrated intensities of the phonon modes \textbf{B}$^2_{\textbf{1g}}$, \textbf{A}$^3_{\textbf{g}}$, and \textbf{A}$^4_{\textbf{g}}$ measured on GeS at $T$=300~K under 2.54~eV, 2.41~eV, and 2.21~eV excitation. The green lines on polar plots are along polarization axes of modes.\label{fig6}}
 \end{figure}  
 
As varying assignments of Raman peaks in GeS were previously reported,\cite{Ho2016,Tan2017,Lam2018,Ribeiro2019} we measured the polarization-resolved RS at $T$=300~K under 1.96~eV excitation. Figure~\ref{fig5} presents polar plots of the integrated intensities as a function of detection angle for all observed phonon modes in co-linear configuration (XX). As the corresponding results in cross-linear configuration (XY) do not add additional value, we focused on XX configuration in our analysis.\cite{ribeiro2015} Solid lines represent fits of the modes intensities as a function of light polarization, I($\theta$), described by:\cite{ribeiro2015}
\begin{equation}
I(\theta)=(|a|sin^2(\theta-\phi)+|c|cos\xi cos^2(\theta-\phi))^2+|c|^2sin^2\xi cos^4(\theta-\phi),
\end{equation}
where $|a|$ and $|c|$ are the amplitudes of the phonon modes, $\phi$ represents the phase of polarization dependence, $\xi$ represents the phase difference. 
It is seen that the polarization axes of the A$^1_{\textrm{g}}$, A$^2_{\textrm{g}}$ and A$^4_{\textrm{g}}$ modes, marked by green lines in the Figure~\ref{fig5}, are approximately oriented in the same direction, $i.e.$ 169$^o$, 168$^o$ and 169$^o$, respectively. The direction is the same as the orientation  apparent in PL and RC spectra (167$^o$ and 170$^o$, respectively), which corresponds to the armchair crystallographic direction. The A$^3_{\textrm{g}}$ mode exhibits different polarization axis of about 112$^o$, which does not match any crystallographic direction. In contrast, the polarization axes of the B$^1_{\textrm{1g}}$ and B$^2_{\textrm{1g}}$ point to the same direction (123$^o$), which is shifted of about 45$^o$ from the crystallographic directions. In terms of observed symmetries of phonon modes, the A$^2_{\textrm{g}}$ and A$^3_{\textrm{g}}$ modes display 2-fold symmetry with an angle period of 180$^o$. In contrast, the A$^{1/4}_{\textrm{g}}$ (B$^{1/2}_{\textrm{1g}}$) mode presents the 4-fold symmetry with an angle period of 90$^o$ and with the different (the same) intensity of perpendicular arms. Consequently, the polarization axes of the other phonon modes (except the A$^3_{\textrm{g}}$) can be used to determine the crystallographic direction, but without the discrimination between the zigzag and armchair directions. Only the 2-fold symmetry of the A$^2_{\textrm{g}}$ mode allows to determine the armchair direction of the crystal.

In order to examine the effect of the excitation energy on the polarization properties of phonon modes, we performed the polarization-resolved RS experiments under three more different excitations (2.21~eV, 2.41~eV and 2.54~eV). 
Due to our experimental limitations, the polarization properties of only three modes, $i.e.$ B$^2_\textrm{{1g}}$, A$^3_{\textrm{g}}$ and A$^4_{\textrm{g}}$, were analyzed, see Figure~\ref{fig6}. Three polarization characteristics of phonon modes can be distinguished: 
(i) The B$^1_{\textrm{g}}$ mode conserves the polarization axis and shape under excitation with different laser wavelength; (ii) The A$^3_{\textrm{g}}$ mode dramatically changes its polarization axis (compare Figures~\ref{fig5} and \ref{fig6}); Moreover, its symmetry also changes: from 2-fold symmetry under 1.96~eV excitation to 4-fold symmetry under other excitations; (iii) In contrast, the 4-fold symmetry of the A$^4_{\textrm{g}}$ mode under 1.96~eV excitation changes gradually to the 2-fold one under 2.41~eV and 2.54~eV excitations. Its 2-fold symmetry can be useful to determine crystallographic direction of the crystal. One can conclude that the different excitation energies may affect significantly the shape between 2- and 4-fold of the A$^3_{\textrm{g}}$ and A$^4_{\textrm{g}}$ modes.

\begin{figure}[h]	
\centering
\includegraphics[width=0.305\linewidth]{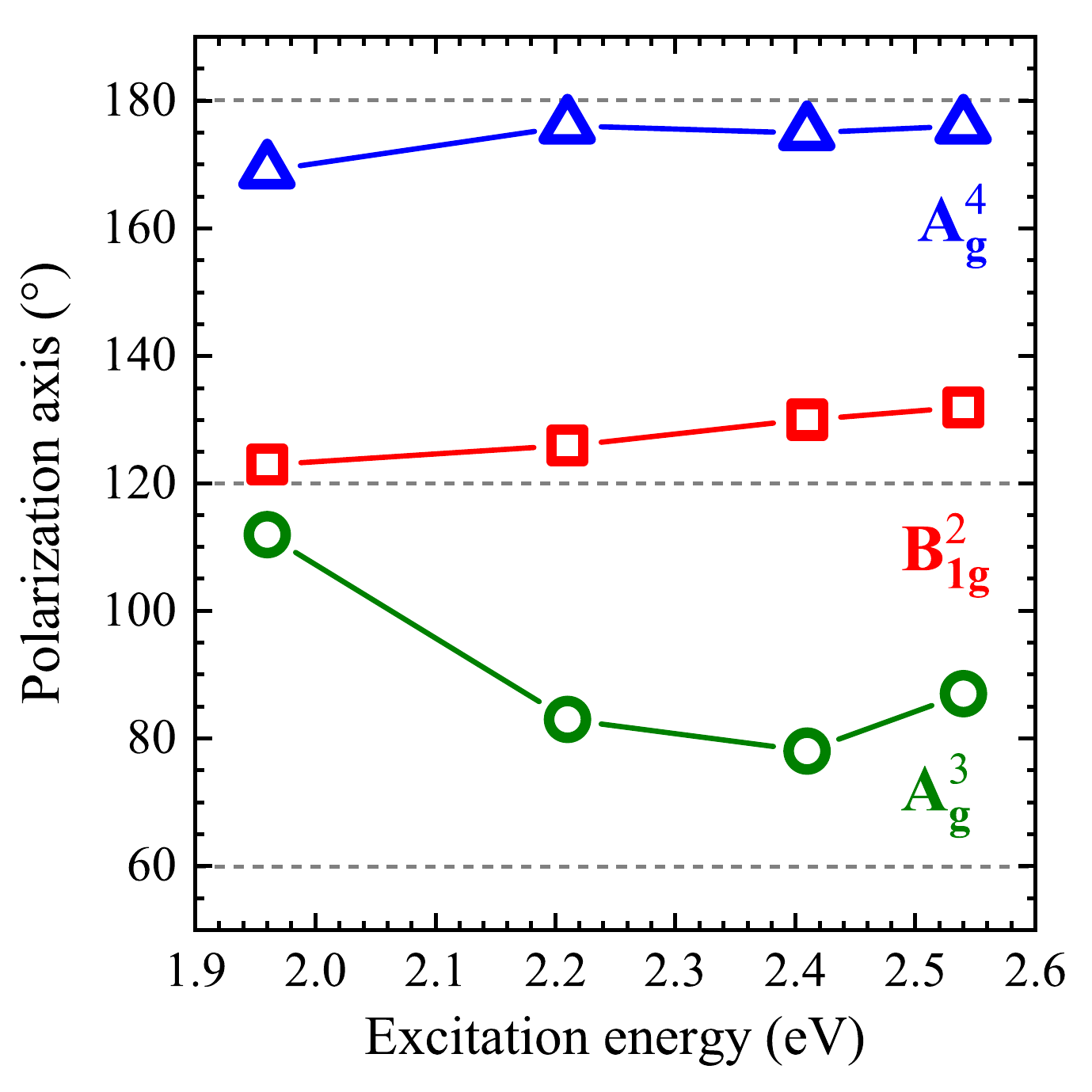}
\caption{The effect of the excitation energy on the polarization axes of three phonon modes: B$^2_\textrm{{1g}}$, A$^3_{\textrm{g}}$ and A$^4_{\textrm{g}}$. 
\label{fig7}}
\end{figure}   

The influence of the excitation energies on the polarization axes of three investigated phonon modes, $i.e.$ B$^2_\textrm{{1g}}$, A$^3_{\textrm{g}}$ and A$^4_{\textrm{g}}$, is summarized in Figure~\ref{fig7}. It can be seen that the polarization axes of the B$^2_\textrm{{1g}}$ and A$^4_{\textrm{g}}$ modes do not change significantly as a function of excitation energy, whereas the difference between the polarization axes of the A$^3_{\textrm{g}}$ mode increases of about 34$^{\circ}$ between the 1.93~eV and 2.41~eV excitations. Due to the observed behavior of different peaks, we can assume that: (i) the polarization properties of the B$^2_\textrm{{1g}}$ mode can be used to determine the crystallographic axes in GeS, but without attribution the zigzag and armchair directions; (ii) the variation of the polarization axis of the A$^3_{\textrm{g}}$ as a function of the excitation energy suggests that the electron-phonon coupling may change the A$^3_{\textrm{g}}$  polarization axis (see Ref. \citenum{Zou2021} for details); (iii) due to 2-fold symmetries of the polarization properties (except for 1.96 eV) and the almost fixed polarization axis of the A$^4_{\textrm{g}}$ mode, its polarization properties are good to identify the zigzag and armchair crystallographic directions. 

It is important to mention that the observed influence of the excitation energies on the axes and shape of polarization properties of phonon modes is very similar to those reported for different anisotropic layered materials, $e.g.$ BP, ReS$_2$, ReSe$_2$, SnSe$_{1-x}$S$_x$.\cite{Kim2015,Xi2016,Choi2020,Sriv2020,Zou2021}
Moreover, as the effect of thickness on the polarization properties of different modes in BP has been reported,\cite{Kim2015,Xi2016} the outline of further research on thin layers of GeS seems to be clear.


\section{Conclusions}

We have presented systematic studies of the optical and vibrational properties of GeS.  It has been found that while the low-temperature ($T$=5 K) optical band-gap absorption is governed by a single resonance related to the neutral exciton, the corresponding emission is dominated by the disorder/impurity- and/or phonon-assisted recombination processes. Moreover, both the RC and PL spectra are found to be linearly polarized along the armchair crystallographic direction. It is proposed to use the effect to determine crystallographic direction of GeS. The effect of the excitation energy on the polarization properties of different phonon modes has been analyzed. It has been shown that the polarization orientation of the A$^2_{\textrm{g}}$ and A$^4_{\textrm{g}}$ modes under specific excitation energy can be useful tools to determine the GeS crystallographic directions: armchair and zigzag. The strong dependence of the  A$^3_{\textrm{g}}$ mode polarization on the excitation light energy strongly suggests its strong coupling to electronic excitation of the crystal.  We believe that the observations will trigger more theoretical studies to explain the origin of the electron-phonon interaction.

\section*{Acknowledgements}
The work has been supported by the National Science Centre, Poland (Grant No. 2017/27/B/ST3/00205 and 2018/31/B/ST3/02111).

\bibliographystyle{apsrev4-1}
\bibliography{biblio}

\begin{thebibliography}{32}%
\makeatletter
\providecommand \@ifxundefined [1]{%
 \@ifx{#1\undefined}
}%
\providecommand \@ifnum [1]{%
 \ifnum #1\expandafter \@firstoftwo
 \else \expandafter \@secondoftwo
 \fi
}%
\providecommand \@ifx [1]{%
 \ifx #1\expandafter \@firstoftwo
 \else \expandafter \@secondoftwo
 \fi
}%
\providecommand \natexlab [1]{#1}%
\providecommand \enquote  [1]{``#1''}%
\providecommand \bibnamefont  [1]{#1}%
\providecommand \bibfnamefont [1]{#1}%
\providecommand \citenamefont [1]{#1}%
\providecommand \href@noop [0]{\@secondoftwo}%
\providecommand \href [0]{\begingroup \@sanitize@url \@href}%
\providecommand \@href[1]{\@@startlink{#1}\@@href}%
\providecommand \@@href[1]{\endgroup#1\@@endlink}%
\providecommand \@sanitize@url [0]{\catcode `\\12\catcode `\$12\catcode
  `\&12\catcode `\#12\catcode `\^12\catcode `\_12\catcode `\%12\relax}%
\providecommand \@@startlink[1]{}%
\providecommand \@@endlink[0]{}%
\providecommand \url  [0]{\begingroup\@sanitize@url \@url }%
\providecommand \@url [1]{\endgroup\@href {#1}{\urlprefix }}%
\providecommand \urlprefix  [0]{URL }%
\providecommand \Eprint [0]{\href }%
\providecommand \doibase [0]{http://dx.doi.org/}%
\providecommand \selectlanguage [0]{\@gobble}%
\providecommand \bibinfo  [0]{\@secondoftwo}%
\providecommand \bibfield  [0]{\@secondoftwo}%
\providecommand \translation [1]{[#1]}%
\providecommand \BibitemOpen [0]{}%
\providecommand \bibitemStop [0]{}%
\providecommand \bibitemNoStop [0]{.\EOS\space}%
\providecommand \EOS [0]{\spacefactor3000\relax}%
\providecommand \BibitemShut  [1]{\csname bibitem#1\endcsname}%
\let\auto@bib@innerbib\@empty
\bibitem [{\citenamefont {Mak}\ \emph {et~al.}(2010)\citenamefont {Mak},
  \citenamefont {Lee}, \citenamefont {Hone}, \citenamefont {Shan},\ and\
  \citenamefont {Heinz}}]{mak2010}%
  \BibitemOpen
  \bibfield  {author} {\bibinfo {author} {\bibfnamefont {K.~F.}\ \bibnamefont
  {Mak}}, \bibinfo {author} {\bibfnamefont {C.}~\bibnamefont {Lee}}, \bibinfo
  {author} {\bibfnamefont {J.}~\bibnamefont {Hone}}, \bibinfo {author}
  {\bibfnamefont {J.}~\bibnamefont {Shan}}, \ and\ \bibinfo {author}
  {\bibfnamefont {T.~F.}\ \bibnamefont {Heinz}},\ }\href {\doibase
  10.1103/PhysRevLett.105.136805} {\bibfield  {journal} {\bibinfo  {journal}
  {Phys. Rev. Lett.}\ }\textbf {\bibinfo {volume} {105}},\ \bibinfo {pages}
  {136805} (\bibinfo {year} {2010})}\BibitemShut {NoStop}%
\bibitem [{\citenamefont {Arora}\ \emph
  {et~al.}(2015{\natexlab{a}})\citenamefont {Arora}, \citenamefont {Koperski},
  \citenamefont {Nogajewski}, \citenamefont {Marcus}, \citenamefont
  {Faugeras},\ and\ \citenamefont {Potemski}}]{AroraWSe2}%
  \BibitemOpen
  \bibfield  {author} {\bibinfo {author} {\bibfnamefont {A.}~\bibnamefont
  {Arora}}, \bibinfo {author} {\bibfnamefont {M.}~\bibnamefont {Koperski}},
  \bibinfo {author} {\bibfnamefont {K.}~\bibnamefont {Nogajewski}}, \bibinfo
  {author} {\bibfnamefont {J.}~\bibnamefont {Marcus}}, \bibinfo {author}
  {\bibfnamefont {C.}~\bibnamefont {Faugeras}}, \ and\ \bibinfo {author}
  {\bibfnamefont {M.}~\bibnamefont {Potemski}},\ }\href {\doibase
  10.1039/C5NR01536G} {\bibfield  {journal} {\bibinfo  {journal} {Nanoscale}\
  }\textbf {\bibinfo {volume} {7}},\ \bibinfo {pages} {10421} (\bibinfo {year}
  {2015}{\natexlab{a}})}\BibitemShut {NoStop}%
\bibitem [{\citenamefont {Arora}\ \emph
  {et~al.}(2015{\natexlab{b}})\citenamefont {Arora}, \citenamefont
  {Nogajewski}, \citenamefont {Molas}, \citenamefont {Koperski},\ and\
  \citenamefont {Potemski}}]{AroraMoSe2}%
  \BibitemOpen
  \bibfield  {author} {\bibinfo {author} {\bibfnamefont {A.}~\bibnamefont
  {Arora}}, \bibinfo {author} {\bibfnamefont {K.}~\bibnamefont {Nogajewski}},
  \bibinfo {author} {\bibfnamefont {M.}~\bibnamefont {Molas}}, \bibinfo
  {author} {\bibfnamefont {M.}~\bibnamefont {Koperski}}, \ and\ \bibinfo
  {author} {\bibfnamefont {M.}~\bibnamefont {Potemski}},\ }\href {\doibase
  10.1039/C5NR06782K} {\bibfield  {journal} {\bibinfo  {journal} {Nanoscale}\
  }\textbf {\bibinfo {volume} {7}},\ \bibinfo {pages} {20769} (\bibinfo {year}
  {2015}{\natexlab{b}})}\BibitemShut {NoStop}%
\bibitem [{\citenamefont {Molas}\ \emph {et~al.}(2017)\citenamefont {Molas},
  \citenamefont {Nogajewski}, \citenamefont {Slobodeniuk}, \citenamefont
  {Binder}, \citenamefont {Bartos},\ and\ \citenamefont {Potemski}}]{MolasWS2}%
  \BibitemOpen
  \bibfield  {author} {\bibinfo {author} {\bibfnamefont {M.~R.}\ \bibnamefont
  {Molas}}, \bibinfo {author} {\bibfnamefont {K.}~\bibnamefont {Nogajewski}},
  \bibinfo {author} {\bibfnamefont {A.~O.}\ \bibnamefont {Slobodeniuk}},
  \bibinfo {author} {\bibfnamefont {J.}~\bibnamefont {Binder}}, \bibinfo
  {author} {\bibfnamefont {M.}~\bibnamefont {Bartos}}, \ and\ \bibinfo {author}
  {\bibfnamefont {M.}~\bibnamefont {Potemski}},\ }\href {\doibase
  10.1039/C7NR04672C} {\bibfield  {journal} {\bibinfo  {journal} {Nanoscale}\
  }\textbf {\bibinfo {volume} {9}},\ \bibinfo {pages} {13128} (\bibinfo {year}
  {2017})}\BibitemShut {NoStop}%
\bibitem [{\citenamefont {Ling}\ \emph
  {et~al.}(2016{\natexlab{a}})\citenamefont {Ling}, , \citenamefont {Huang},
  \citenamefont {Hasdeo}, \citenamefont {Liang}, \citenamefont {Parkin},
  \citenamefont {Tatsumi}, \citenamefont {Nugraha}, \citenamefont {Puretzky},
  \citenamefont {Das}, \citenamefont {Sumpter}, \citenamefont {Geohegan},
  \citenamefont {Kong}, \citenamefont {Saito}, \citenamefont {Drndic},
  \citenamefont {Meunier},\ and\ \citenamefont {Dresselhaus}}]{Ling2016}%
  \BibitemOpen
  \bibfield  {author} {\bibinfo {author} {\bibfnamefont {X.}~\bibnamefont
  {Ling}}, , \bibinfo {author} {\bibfnamefont {S.}~\bibnamefont {Huang}},
  \bibinfo {author} {\bibfnamefont {E.~H.}\ \bibnamefont {Hasdeo}}, \bibinfo
  {author} {\bibfnamefont {L.}~\bibnamefont {Liang}}, \bibinfo {author}
  {\bibfnamefont {W.~M.}\ \bibnamefont {Parkin}}, \bibinfo {author}
  {\bibfnamefont {Y.}~\bibnamefont {Tatsumi}}, \bibinfo {author} {\bibfnamefont
  {A.~R.~T.}\ \bibnamefont {Nugraha}}, \bibinfo {author} {\bibfnamefont
  {A.~A.}\ \bibnamefont {Puretzky}}, \bibinfo {author} {\bibfnamefont {P.~M.}\
  \bibnamefont {Das}}, \bibinfo {author} {\bibfnamefont {B.~G.}\ \bibnamefont
  {Sumpter}}, \bibinfo {author} {\bibfnamefont {D.~B.}\ \bibnamefont
  {Geohegan}}, \bibinfo {author} {\bibfnamefont {J.}~\bibnamefont {Kong}},
  \bibinfo {author} {\bibfnamefont {R.}~\bibnamefont {Saito}}, \bibinfo
  {author} {\bibfnamefont {M.}~\bibnamefont {Drndic}}, \bibinfo {author}
  {\bibfnamefont {V.}~\bibnamefont {Meunier}}, \ and\ \bibinfo {author}
  {\bibfnamefont {M.~S.}\ \bibnamefont {Dresselhaus}},\ }\href {\doibase
  10.1021/acs.nanolett.5b04540} {\bibfield  {journal} {\bibinfo  {journal}
  {Nano Letters}\ }\textbf {\bibinfo {volume} {16}},\ \bibinfo {pages} {2260}
  (\bibinfo {year} {2016}{\natexlab{a}})}\BibitemShut {NoStop}%
\bibitem [{\citenamefont {Ribeiro}\ \emph {et~al.}(2015)\citenamefont
  {Ribeiro}, \citenamefont {Pimenta}, \citenamefont {de~Matos}, \citenamefont
  {Moreira}, \citenamefont {Rodin}, \citenamefont {Zapata}, \citenamefont
  {de~Souza},\ and\ \citenamefont {Neto}}]{ribeiro2015}%
  \BibitemOpen
  \bibfield  {author} {\bibinfo {author} {\bibfnamefont {H.~B.}\ \bibnamefont
  {Ribeiro}}, \bibinfo {author} {\bibfnamefont {M.~A.}\ \bibnamefont
  {Pimenta}}, \bibinfo {author} {\bibfnamefont {C.~J.~S.}\ \bibnamefont
  {de~Matos}}, \bibinfo {author} {\bibfnamefont {R.~L.}\ \bibnamefont
  {Moreira}}, \bibinfo {author} {\bibfnamefont {A.~S.}\ \bibnamefont {Rodin}},
  \bibinfo {author} {\bibfnamefont {J.~D.}\ \bibnamefont {Zapata}}, \bibinfo
  {author} {\bibfnamefont {E.~A.~T.}\ \bibnamefont {de~Souza}}, \ and\ \bibinfo
  {author} {\bibfnamefont {A.~H.~C.}\ \bibnamefont {Neto}},\ }\href {\doibase
  10.1021/acsnano.5b00698} {\bibfield  {journal} {\bibinfo  {journal} {ACS
  Nano}\ }\textbf {\bibinfo {volume} {9}},\ \bibinfo {pages} {4270} (\bibinfo
  {year} {2015})}\BibitemShut {NoStop}%
\bibitem [{\citenamefont {Molas}\ \emph {et~al.}(2021)\citenamefont {Molas},
  \citenamefont {Macewicz}, \citenamefont {Wieloszy\'nska}, \citenamefont
  {Jak\'obczyk}, \citenamefont {Wysmo\l{}ek}, \citenamefont {Bogdanowicz},\
  and\ \citenamefont {Jasinski}}]{molas2021}%
  \BibitemOpen
  \bibfield  {author} {\bibinfo {author} {\bibfnamefont {M.~R.}\ \bibnamefont
  {Molas}}, \bibinfo {author} {\bibfnamefont {L.}~\bibnamefont {Macewicz}},
  \bibinfo {author} {\bibfnamefont {A.}~\bibnamefont {Wieloszy\'nska}},
  \bibinfo {author} {\bibfnamefont {P.}~\bibnamefont {Jak\'obczyk}}, \bibinfo
  {author} {\bibfnamefont {A.}~\bibnamefont {Wysmo\l{}ek}}, \bibinfo {author}
  {\bibfnamefont {R.}~\bibnamefont {Bogdanowicz}}, \ and\ \bibinfo {author}
  {\bibfnamefont {J.~B.}\ \bibnamefont {Jasinski}},\ }\href {\doibase
  10.1038/s41699-021-00263-8} {\bibfield  {journal} {\bibinfo  {journal} {npj
  2D Materials and Applications}\ }\textbf {\bibinfo {volume} {5}} (\bibinfo
  {year} {2021}),\ 10.1038/s41699-021-00263-8}\BibitemShut {NoStop}%
\bibitem [{\citenamefont {Li}\ \emph {et~al.}(2016)\citenamefont {Li},
  \citenamefont {Liu}, \citenamefont {Wang},\ and\ \citenamefont
  {Li}}]{Li2016}%
  \BibitemOpen
  \bibfield  {author} {\bibinfo {author} {\bibfnamefont {F.}~\bibnamefont
  {Li}}, \bibinfo {author} {\bibfnamefont {X.}~\bibnamefont {Liu}}, \bibinfo
  {author} {\bibfnamefont {Y.}~\bibnamefont {Wang}}, \ and\ \bibinfo {author}
  {\bibfnamefont {Y.}~\bibnamefont {Li}},\ }\href {\doibase 10.1039/C6TC00454G}
  {\bibfield  {journal} {\bibinfo  {journal} {Journal of Materials Chemistry
  C}\ }\textbf {\bibinfo {volume} {4}},\ \bibinfo {pages} {2155} (\bibinfo
  {year} {2016})}\BibitemShut {NoStop}%
\bibitem [{\citenamefont {Kresse}\ and\ \citenamefont
  {Furthm\"uller}(1996)}]{Kresse1996}%
  \BibitemOpen
  \bibfield  {author} {\bibinfo {author} {\bibfnamefont {G.}~\bibnamefont
  {Kresse}}\ and\ \bibinfo {author} {\bibfnamefont {J.}~\bibnamefont
  {Furthm\"uller}},\ }\href {\doibase 10.1103/PhysRevB.54.11169} {\bibfield
  {journal} {\bibinfo  {journal} {Phys. Rev. B}\ }\textbf {\bibinfo {volume}
  {54}},\ \bibinfo {pages} {11169} (\bibinfo {year} {1996})}\BibitemShut
  {NoStop}%
\bibitem [{\citenamefont {Kresse}\ and\ \citenamefont
  {Joubert}(1999)}]{Kresse1999}%
  \BibitemOpen
  \bibfield  {author} {\bibinfo {author} {\bibfnamefont {G.}~\bibnamefont
  {Kresse}}\ and\ \bibinfo {author} {\bibfnamefont {D.}~\bibnamefont
  {Joubert}},\ }\href {\doibase 10.1103/PhysRevB.59.1758} {\bibfield  {journal}
  {\bibinfo  {journal} {Phys. Rev. B}\ }\textbf {\bibinfo {volume} {59}},\
  \bibinfo {pages} {1758} (\bibinfo {year} {1999})}\BibitemShut {NoStop}%
\bibitem [{\citenamefont {Perdew}\ \emph {et~al.}(1996)\citenamefont {Perdew},
  \citenamefont {Burke},\ and\ \citenamefont {Ernzerhof}}]{Perdew1996}%
  \BibitemOpen
  \bibfield  {author} {\bibinfo {author} {\bibfnamefont {J.~P.}\ \bibnamefont
  {Perdew}}, \bibinfo {author} {\bibfnamefont {K.}~\bibnamefont {Burke}}, \
  and\ \bibinfo {author} {\bibfnamefont {M.}~\bibnamefont {Ernzerhof}},\ }\href
  {\doibase 10.1103/PhysRevLett.77.3865} {\bibfield  {journal} {\bibinfo
  {journal} {Phys. Rev. Lett.}\ }\textbf {\bibinfo {volume} {77}},\ \bibinfo
  {pages} {3865} (\bibinfo {year} {1996})}\BibitemShut {NoStop}%
\bibitem [{\citenamefont {Grimme}\ \emph {et~al.}(2010)\citenamefont {Grimme},
  \citenamefont {Antony}, \citenamefont {Ehrlich},\ and\ \citenamefont
  {Krieg}}]{Grimme2010}%
  \BibitemOpen
  \bibfield  {author} {\bibinfo {author} {\bibfnamefont {S.}~\bibnamefont
  {Grimme}}, \bibinfo {author} {\bibfnamefont {J.}~\bibnamefont {Antony}},
  \bibinfo {author} {\bibfnamefont {S.}~\bibnamefont {Ehrlich}}, \ and\
  \bibinfo {author} {\bibfnamefont {H.}~\bibnamefont {Krieg}},\ }\href
  {\doibase 10.1063/1.3382344} {\bibfield  {journal} {\bibinfo  {journal} {The
  Journal of Chemical Physics}\ }\textbf {\bibinfo {volume} {132}},\ \bibinfo
  {pages} {154104} (\bibinfo {year} {2010})}\BibitemShut {NoStop}%
\bibitem [{\citenamefont {Parlinski}\ \emph {et~al.}(1997)\citenamefont
  {Parlinski}, \citenamefont {Li},\ and\ \citenamefont
  {Kawazoe}}]{Parlinski1997}%
  \BibitemOpen
  \bibfield  {author} {\bibinfo {author} {\bibfnamefont {K.}~\bibnamefont
  {Parlinski}}, \bibinfo {author} {\bibfnamefont {Z.~Q.}\ \bibnamefont {Li}}, \
  and\ \bibinfo {author} {\bibfnamefont {Y.}~\bibnamefont {Kawazoe}},\ }\href
  {\doibase 10.1103/PhysRevLett.78.4063} {\bibfield  {journal} {\bibinfo
  {journal} {Phys. Rev. Lett.}\ }\textbf {\bibinfo {volume} {78}},\ \bibinfo
  {pages} {4063} (\bibinfo {year} {1997})}\BibitemShut {NoStop}%
\bibitem [{\citenamefont {Togo}\ and\ \citenamefont {Tanaka}(2015)}]{Togo2015}%
  \BibitemOpen
  \bibfield  {author} {\bibinfo {author} {\bibfnamefont {A.}~\bibnamefont
  {Togo}}\ and\ \bibinfo {author} {\bibfnamefont {I.}~\bibnamefont {Tanaka}},\
  }\href {\doibase 10.1016/j.scriptamat.2015.07.021} {\bibfield  {journal}
  {\bibinfo  {journal} {Scripta Materialia}\ }\textbf {\bibinfo {volume}
  {108}},\ \bibinfo {pages} {1} (\bibinfo {year} {2015})}\BibitemShut {NoStop}%
\bibitem [{\citenamefont {Togo}\ and\ \citenamefont {Tanaka}(2018)}]{Togo2018}%
  \BibitemOpen
  \bibfield  {author} {\bibinfo {author} {\bibfnamefont {A.}~\bibnamefont
  {Togo}}\ and\ \bibinfo {author} {\bibfnamefont {I.}~\bibnamefont {Tanaka}},\
  }\href@noop {} {\  (\bibinfo {year} {2018})},\ \Eprint
  {http://arxiv.org/abs/1808.01590} {arXiv:1808.01590 [cond-mat.mtrl-sci]}
  \BibitemShut {NoStop}%
\bibitem [{\citenamefont {Wiedemeier}\ and\ \citenamefont
  {Siemers}(1977)}]{Siemers}%
  \BibitemOpen
  \bibfield  {author} {\bibinfo {author} {\bibfnamefont {H.}~\bibnamefont
  {Wiedemeier}}\ and\ \bibinfo {author} {\bibfnamefont {P.~A.}\ \bibnamefont
  {Siemers}},\ }\href {\doibase 10.1002/zaac.19774310134} {\bibfield  {journal}
  {\bibinfo  {journal} {Zeitschrift für anorganische und allgemeine Chemie}\
  }\textbf {\bibinfo {volume} {431}},\ \bibinfo {pages} {299} (\bibinfo {year}
  {1977})}\BibitemShut {NoStop}%
\bibitem [{\citenamefont {Hecht}(2017)}]{Hecht}%
  \BibitemOpen
  \bibfield  {author} {\bibinfo {author} {\bibfnamefont {E.}~\bibnamefont
  {Hecht}},\ }\href@noop {} {\emph {\bibinfo {title} {OPTics, ed. 5$^th$}}}\
  (\bibinfo  {publisher} {Pearson Education Limited},\ \bibinfo {year} {2017})\
  p.\ \bibinfo {pages} {347}\BibitemShut {NoStop}%
\bibitem [{\citenamefont {Ho}\ \emph {et~al.}(2016)\citenamefont {Ho}, ,\ and\
  \citenamefont {Li}}]{Ho2016}%
  \BibitemOpen
  \bibfield  {author} {\bibinfo {author} {\bibfnamefont {C.~H.}\ \bibnamefont
  {Ho}}, , \ and\ \bibinfo {author} {\bibfnamefont {J.~X.}\ \bibnamefont
  {Li}},\ }\href {\doibase 10.1002/adom.201600814} {\bibfield  {journal}
  {\bibinfo  {journal} {Material Views}\ }\textbf {\bibinfo {volume} {5}},\
  \bibinfo {pages} {1600814} (\bibinfo {year} {2016})}\BibitemShut {NoStop}%
\bibitem [{\citenamefont {Oliva}\ \emph {et~al.}(2020)\citenamefont {Oliva},
  \citenamefont {Wo\ifmmode~\acute{z}\else \'{z}\fi{}niak}, \citenamefont
  {Dybala}, \citenamefont {To\l{}\l{}oczko}, \citenamefont {Kopaczek},
  \citenamefont {Scharoch},\ and\ \citenamefont {Kudrawiec}}]{Oliva2020}%
  \BibitemOpen
  \bibfield  {author} {\bibinfo {author} {\bibfnamefont {R.}~\bibnamefont
  {Oliva}}, \bibinfo {author} {\bibfnamefont {T.}~\bibnamefont
  {Wo\ifmmode~\acute{z}\else \'{z}\fi{}niak}}, \bibinfo {author} {\bibfnamefont
  {F.}~\bibnamefont {Dybala}}, \bibinfo {author} {\bibfnamefont
  {A.}~\bibnamefont {To\l{}\l{}oczko}}, \bibinfo {author} {\bibfnamefont
  {J.}~\bibnamefont {Kopaczek}}, \bibinfo {author} {\bibfnamefont
  {P.}~\bibnamefont {Scharoch}}, \ and\ \bibinfo {author} {\bibfnamefont
  {R.}~\bibnamefont {Kudrawiec}},\ }\href {\doibase
  10.1103/PhysRevB.101.235205} {\bibfield  {journal} {\bibinfo  {journal}
  {Phys. Rev. B}\ }\textbf {\bibinfo {volume} {101}},\ \bibinfo {pages}
  {235205} (\bibinfo {year} {2020})}\BibitemShut {NoStop}%
\bibitem [{\citenamefont {Tołłoczko}\ \emph {et~al.}(2020)\citenamefont
  {Tołłoczko}, \citenamefont {Oliva}, \citenamefont {Wożniak}, \citenamefont
  {Kopaczek}, \citenamefont {Scharoch},\ and\ \citenamefont
  {Kudrawiec}}]{tolloczko2020}%
  \BibitemOpen
  \bibfield  {author} {\bibinfo {author} {\bibfnamefont {A.}~\bibnamefont
  {Tołłoczko}}, \bibinfo {author} {\bibfnamefont {R.}~\bibnamefont {Oliva}},
  \bibinfo {author} {\bibfnamefont {T.}~\bibnamefont {Wożniak}}, \bibinfo
  {author} {\bibfnamefont {J.}~\bibnamefont {Kopaczek}}, \bibinfo {author}
  {\bibfnamefont {P.}~\bibnamefont {Scharoch}}, \ and\ \bibinfo {author}
  {\bibfnamefont {R.}~\bibnamefont {Kudrawiec}},\ }\href {\doibase
  10.1039/d0ma00146e} {\bibfield  {journal} {\bibinfo  {journal} {Material
  Advances}\ }\textbf {\bibinfo {volume} {1}},\ \bibinfo {pages} {1886}
  (\bibinfo {year} {2020})}\BibitemShut {NoStop}%
\bibitem [{\citenamefont {Plechinger}\ \emph {et~al.}(2015)\citenamefont
  {Plechinger}, \citenamefont {Nagler}, \citenamefont {Kraus}, \citenamefont
  {Paradiso}, \citenamefont {Strunk}, \citenamefont {Schüller},\ and\
  \citenamefont {Korn}}]{Plechinger2015}%
  \BibitemOpen
  \bibfield  {author} {\bibinfo {author} {\bibfnamefont {G.}~\bibnamefont
  {Plechinger}}, \bibinfo {author} {\bibfnamefont {P.}~\bibnamefont {Nagler}},
  \bibinfo {author} {\bibfnamefont {J.}~\bibnamefont {Kraus}}, \bibinfo
  {author} {\bibfnamefont {N.}~\bibnamefont {Paradiso}}, \bibinfo {author}
  {\bibfnamefont {C.}~\bibnamefont {Strunk}}, \bibinfo {author} {\bibfnamefont
  {C.}~\bibnamefont {Schüller}}, \ and\ \bibinfo {author} {\bibfnamefont
  {T.}~\bibnamefont {Korn}},\ }\href {\doibase 10.1002/pssr.201510224}
  {\bibfield  {journal} {\bibinfo  {journal} {physica status solidi (RRL) –
  Rapid Research Letters}\ }\textbf {\bibinfo {volume} {9}},\ \bibinfo {pages}
  {457} (\bibinfo {year} {2015})}\BibitemShut {NoStop}%
\bibitem [{\citenamefont {Shang}\ \emph {et~al.}(2015)\citenamefont {Shang},
  \citenamefont {Shen}, \citenamefont {Cong}, \citenamefont {Peimyoo},
  \citenamefont {Cao}, \citenamefont {Eginligil},\ and\ \citenamefont
  {Yu}}]{Shang2015}%
  \BibitemOpen
  \bibfield  {author} {\bibinfo {author} {\bibfnamefont {J.}~\bibnamefont
  {Shang}}, \bibinfo {author} {\bibfnamefont {X.}~\bibnamefont {Shen}},
  \bibinfo {author} {\bibfnamefont {C.}~\bibnamefont {Cong}}, \bibinfo {author}
  {\bibfnamefont {N.}~\bibnamefont {Peimyoo}}, \bibinfo {author} {\bibfnamefont
  {B.}~\bibnamefont {Cao}}, \bibinfo {author} {\bibfnamefont {M.}~\bibnamefont
  {Eginligil}}, \ and\ \bibinfo {author} {\bibfnamefont {T.}~\bibnamefont
  {Yu}},\ }\href {\doibase 10.1021/nn5059908} {\bibfield  {journal} {\bibinfo
  {journal} {ACS Nano}\ }\textbf {\bibinfo {volume} {9}},\ \bibinfo {pages}
  {647} (\bibinfo {year} {2015})}\BibitemShut {NoStop}%
\bibitem [{\citenamefont {Plechinger}\ \emph {et~al.}(2016)\citenamefont
  {Plechinger}, \citenamefont {Nagler}, \citenamefont {Arora}, \citenamefont
  {Granados~del Águila}, \citenamefont {Ballottin}, \citenamefont {Frank},
  \citenamefont {Steinleitner}, \citenamefont {Gmitra}, \citenamefont {Fabian},
  \citenamefont {Christianen}, \citenamefont {Bratschitsch}, \citenamefont
  {Schüller},\ and\ \citenamefont {Korn}}]{Plechinger2016}%
  \BibitemOpen
  \bibfield  {author} {\bibinfo {author} {\bibfnamefont {G.}~\bibnamefont
  {Plechinger}}, \bibinfo {author} {\bibfnamefont {P.}~\bibnamefont {Nagler}},
  \bibinfo {author} {\bibfnamefont {A.}~\bibnamefont {Arora}}, \bibinfo
  {author} {\bibfnamefont {A.}~\bibnamefont {Granados~del Águila}}, \bibinfo
  {author} {\bibfnamefont {M.~V.}\ \bibnamefont {Ballottin}}, \bibinfo {author}
  {\bibfnamefont {T.}~\bibnamefont {Frank}}, \bibinfo {author} {\bibfnamefont
  {P.}~\bibnamefont {Steinleitner}}, \bibinfo {author} {\bibfnamefont
  {M.}~\bibnamefont {Gmitra}}, \bibinfo {author} {\bibfnamefont
  {J.}~\bibnamefont {Fabian}}, \bibinfo {author} {\bibfnamefont {P.~C.~M.}\
  \bibnamefont {Christianen}}, \bibinfo {author} {\bibfnamefont
  {R.}~\bibnamefont {Bratschitsch}}, \bibinfo {author} {\bibfnamefont
  {C.}~\bibnamefont {Schüller}}, \ and\ \bibinfo {author} {\bibfnamefont
  {T.}~\bibnamefont {Korn}},\ }\href {\doibase 10.1021/acs.nanolett.6b04171}
  {\bibfield  {journal} {\bibinfo  {journal} {Nano Letters}\ }\textbf {\bibinfo
  {volume} {16}},\ \bibinfo {pages} {7899} (\bibinfo {year}
  {2016})}\BibitemShut {NoStop}%
\bibitem [{\citenamefont {K{\l}opotowski}\ \emph {et~al.}(2016)\citenamefont
  {K{\l}opotowski}, \citenamefont {Backes}, \citenamefont {Mitioglu},
  \citenamefont {Vega-Mayoral}, \citenamefont {Hanlon}, \citenamefont
  {Coleman}, \citenamefont {Ivanov}, \citenamefont {Maude},\ and\ \citenamefont
  {Plochocka}}]{Klopotowski2016}%
  \BibitemOpen
  \bibfield  {author} {\bibinfo {author} {\bibfnamefont {{\L}.}~\bibnamefont
  {K{\l}opotowski}}, \bibinfo {author} {\bibfnamefont {C.}~\bibnamefont
  {Backes}}, \bibinfo {author} {\bibfnamefont {A.~A.}\ \bibnamefont
  {Mitioglu}}, \bibinfo {author} {\bibfnamefont {V.}~\bibnamefont
  {Vega-Mayoral}}, \bibinfo {author} {\bibfnamefont {D.}~\bibnamefont
  {Hanlon}}, \bibinfo {author} {\bibfnamefont {J.~N.}\ \bibnamefont {Coleman}},
  \bibinfo {author} {\bibfnamefont {V.~Y.}\ \bibnamefont {Ivanov}}, \bibinfo
  {author} {\bibfnamefont {D.~K.}\ \bibnamefont {Maude}}, \ and\ \bibinfo
  {author} {\bibfnamefont {P.}~\bibnamefont {Plochocka}},\ }\href {\doibase
  10.1088/0957-4484/27/42/425701} {\bibfield  {journal} {\bibinfo  {journal}
  {Nanotechnology}\ }\textbf {\bibinfo {volume} {27}},\ \bibinfo {pages}
  {425701} (\bibinfo {year} {2016})}\BibitemShut {NoStop}%
\bibitem [{\citenamefont {Tan}\ \emph {et~al.}(2017)\citenamefont {Tan},
  \citenamefont {Lim}, \citenamefont {Wang}, \citenamefont {Mohamed},
  \citenamefont {Moun}, \citenamefont {Zhang}, \citenamefont {Miyauchu},
  \citenamefont {Ohfuchi},\ and\ \citenamefont {Matsuda}}]{Tan2017}%
  \BibitemOpen
  \bibfield  {author} {\bibinfo {author} {\bibfnamefont {D.}~\bibnamefont
  {Tan}}, \bibinfo {author} {\bibfnamefont {H.~E.}\ \bibnamefont {Lim}},
  \bibinfo {author} {\bibfnamefont {F.}~\bibnamefont {Wang}}, \bibinfo {author}
  {\bibfnamefont {N.~B.}\ \bibnamefont {Mohamed}}, \bibinfo {author}
  {\bibfnamefont {S.}~\bibnamefont {Moun}}, \bibinfo {author} {\bibfnamefont
  {W.}~\bibnamefont {Zhang}}, \bibinfo {author} {\bibfnamefont
  {Y.}~\bibnamefont {Miyauchu}}, \bibinfo {author} {\bibfnamefont
  {M.}~\bibnamefont {Ohfuchi}}, \ and\ \bibinfo {author} {\bibfnamefont
  {K.}~\bibnamefont {Matsuda}},\ }\href {\doibase 10.1007/s12274-016-1312-6}
  {\bibfield  {journal} {\bibinfo  {journal} {Nano Research}\ }\textbf
  {\bibinfo {volume} {2}},\ \bibinfo {pages} {546} (\bibinfo {year}
  {2017})}\BibitemShut {NoStop}%
\bibitem [{\citenamefont {Lam}\ \emph {et~al.}(2018)\citenamefont {Lam},
  \citenamefont {Chen}, \citenamefont {Kang}, \citenamefont {Liu},\ and\
  \citenamefont {Hersam}}]{Lam2018}%
  \BibitemOpen
  \bibfield  {author} {\bibinfo {author} {\bibfnamefont {D.}~\bibnamefont
  {Lam}}, \bibinfo {author} {\bibfnamefont {K.-S.}\ \bibnamefont {Chen}},
  \bibinfo {author} {\bibfnamefont {J.}~\bibnamefont {Kang}}, \bibinfo {author}
  {\bibfnamefont {X.}~\bibnamefont {Liu}}, \ and\ \bibinfo {author}
  {\bibfnamefont {M.~C.}\ \bibnamefont {Hersam}},\ }\href {\doibase
  10.1021/acs.chemmater.7b04652} {\bibfield  {journal} {\bibinfo  {journal}
  {Chemistry of Materials}\ }\textbf {\bibinfo {volume} {30}},\ \bibinfo
  {pages} {2245} (\bibinfo {year} {2018})}\BibitemShut {NoStop}%
\bibitem [{\citenamefont {Ribeiro}\ \emph {et~al.}(2019)\citenamefont
  {Ribeiro}, \citenamefont {Ramos}, \citenamefont {Seixas}, \citenamefont
  {de~Matos},\ and\ \citenamefont {Pimenta}}]{Ribeiro2019}%
  \BibitemOpen
  \bibfield  {author} {\bibinfo {author} {\bibfnamefont {H.~B.}\ \bibnamefont
  {Ribeiro}}, \bibinfo {author} {\bibfnamefont {S.~L. L.~M.}\ \bibnamefont
  {Ramos}}, \bibinfo {author} {\bibfnamefont {L.}~\bibnamefont {Seixas}},
  \bibinfo {author} {\bibfnamefont {C.~J.~S.}\ \bibnamefont {de~Matos}}, \ and\
  \bibinfo {author} {\bibfnamefont {M.~A.}\ \bibnamefont {Pimenta}},\ }\href
  {\doibase 10.1103/PhysRevB.100.094301} {\bibfield  {journal} {\bibinfo
  {journal} {Phys. Rev. B}\ }\textbf {\bibinfo {volume} {100}},\ \bibinfo
  {pages} {094301} (\bibinfo {year} {2019})}\BibitemShut {NoStop}%
\bibitem [{\citenamefont {Zou}\ \emph {et~al.}(2021)\citenamefont {Zou},
  \citenamefont {Wei}, \citenamefont {Zhou}, \citenamefont {Ke}, \citenamefont
  {Zhang}, \citenamefont {Zhang}, \citenamefont {Yip}, \citenamefont {Chen},
  \citenamefont {Li},\ and\ \citenamefont {Sun}}]{Zou2021}%
  \BibitemOpen
  \bibfield  {author} {\bibinfo {author} {\bibfnamefont {B.}~\bibnamefont
  {Zou}}, \bibinfo {author} {\bibfnamefont {Y.}~\bibnamefont {Wei}}, \bibinfo
  {author} {\bibfnamefont {Y.}~\bibnamefont {Zhou}}, \bibinfo {author}
  {\bibfnamefont {D.}~\bibnamefont {Ke}}, \bibinfo {author} {\bibfnamefont
  {X.}~\bibnamefont {Zhang}}, \bibinfo {author} {\bibfnamefont
  {M.}~\bibnamefont {Zhang}}, \bibinfo {author} {\bibfnamefont {C.-T.}\
  \bibnamefont {Yip}}, \bibinfo {author} {\bibfnamefont {X.}~\bibnamefont
  {Chen}}, \bibinfo {author} {\bibfnamefont {W.}~\bibnamefont {Li}}, \ and\
  \bibinfo {author} {\bibfnamefont {H.}~\bibnamefont {Sun}},\ }\href {\doibase
  10.1039/D1NH00220A} {\bibfield  {journal} {\bibinfo  {journal} {Nanoscale
  Horiz.}\ }\textbf {\bibinfo {volume} {6}},\ \bibinfo {pages} {809} (\bibinfo
  {year} {2021})}\BibitemShut {NoStop}%
\bibitem [{\citenamefont {Kim}\ \emph {et~al.}(2015)\citenamefont {Kim},
  \citenamefont {Lee}, \citenamefont {Lee}, \citenamefont {Park}, \citenamefont
  {Lee}, \citenamefont {Lee},\ and\ \citenamefont {Cheong}}]{Kim2015}%
  \BibitemOpen
  \bibfield  {author} {\bibinfo {author} {\bibfnamefont {J.}~\bibnamefont
  {Kim}}, \bibinfo {author} {\bibfnamefont {J.-U.}\ \bibnamefont {Lee}},
  \bibinfo {author} {\bibfnamefont {J.}~\bibnamefont {Lee}}, \bibinfo {author}
  {\bibfnamefont {H.~J.}\ \bibnamefont {Park}}, \bibinfo {author}
  {\bibfnamefont {Z.}~\bibnamefont {Lee}}, \bibinfo {author} {\bibfnamefont
  {C.}~\bibnamefont {Lee}}, \ and\ \bibinfo {author} {\bibfnamefont
  {H.}~\bibnamefont {Cheong}},\ }\href {\doibase 10.1039/C5NR04349B} {\bibfield
   {journal} {\bibinfo  {journal} {Nanoscale}\ }\textbf {\bibinfo {volume}
  {7}},\ \bibinfo {pages} {18708} (\bibinfo {year} {2015})}\BibitemShut
  {NoStop}%
\bibitem [{\citenamefont {Ling}\ \emph
  {et~al.}(2016{\natexlab{b}})\citenamefont {Ling}, \citenamefont {Huang},
  \citenamefont {Hasdeo}, \citenamefont {Liang}, \citenamefont {Parkin},
  \citenamefont {Tatsumi}, \citenamefont {Nugraha}, \citenamefont {Puretzky},
  \citenamefont {Das}, \citenamefont {Sumpter}, \citenamefont {Geohegan},
  \citenamefont {Kong}, \citenamefont {Saito}, \citenamefont {Drndic},
  \citenamefont {Meunier},\ and\ \citenamefont {Dresselhaus}}]{Xi2016}%
  \BibitemOpen
  \bibfield  {author} {\bibinfo {author} {\bibfnamefont {X.}~\bibnamefont
  {Ling}}, \bibinfo {author} {\bibfnamefont {S.}~\bibnamefont {Huang}},
  \bibinfo {author} {\bibfnamefont {E.~H.}\ \bibnamefont {Hasdeo}}, \bibinfo
  {author} {\bibfnamefont {L.}~\bibnamefont {Liang}}, \bibinfo {author}
  {\bibfnamefont {W.~M.}\ \bibnamefont {Parkin}}, \bibinfo {author}
  {\bibfnamefont {Y.}~\bibnamefont {Tatsumi}}, \bibinfo {author} {\bibfnamefont
  {A.~R.~T.}\ \bibnamefont {Nugraha}}, \bibinfo {author} {\bibfnamefont
  {A.~A.}\ \bibnamefont {Puretzky}}, \bibinfo {author} {\bibfnamefont {P.~M.}\
  \bibnamefont {Das}}, \bibinfo {author} {\bibfnamefont {B.~G.}\ \bibnamefont
  {Sumpter}}, \bibinfo {author} {\bibfnamefont {D.~B.}\ \bibnamefont
  {Geohegan}}, \bibinfo {author} {\bibfnamefont {J.}~\bibnamefont {Kong}},
  \bibinfo {author} {\bibfnamefont {R.}~\bibnamefont {Saito}}, \bibinfo
  {author} {\bibfnamefont {M.}~\bibnamefont {Drndic}}, \bibinfo {author}
  {\bibfnamefont {V.}~\bibnamefont {Meunier}}, \ and\ \bibinfo {author}
  {\bibfnamefont {M.~S.}\ \bibnamefont {Dresselhaus}},\ }\href {\doibase
  10.1021/acs.nanolett.5b04540} {\bibfield  {journal} {\bibinfo  {journal}
  {Nano Letters}\ }\textbf {\bibinfo {volume} {16}},\ \bibinfo {pages} {2260}
  (\bibinfo {year} {2016}{\natexlab{b}})},\ \bibinfo {note} {pMID:
  26963685}\BibitemShut {NoStop}%
\bibitem [{\citenamefont {Choi}\ \emph {et~al.}(2020)\citenamefont {Choi},
  \citenamefont {Kim}, \citenamefont {Lim}, \citenamefont {Kim}, \citenamefont
  {Park}, \citenamefont {Kim}, \citenamefont {Lee},\ and\ \citenamefont
  {Cheong}}]{Choi2020}%
  \BibitemOpen
  \bibfield  {author} {\bibinfo {author} {\bibfnamefont {Y.}~\bibnamefont
  {Choi}}, \bibinfo {author} {\bibfnamefont {K.}~\bibnamefont {Kim}}, \bibinfo
  {author} {\bibfnamefont {S.~Y.}\ \bibnamefont {Lim}}, \bibinfo {author}
  {\bibfnamefont {J.}~\bibnamefont {Kim}}, \bibinfo {author} {\bibfnamefont
  {J.~M.}\ \bibnamefont {Park}}, \bibinfo {author} {\bibfnamefont {J.~H.}\
  \bibnamefont {Kim}}, \bibinfo {author} {\bibfnamefont {Z.}~\bibnamefont
  {Lee}}, \ and\ \bibinfo {author} {\bibfnamefont {H.}~\bibnamefont {Cheong}},\
  }\href {\doibase 10.1039/C9NH00487D} {\bibfield  {journal} {\bibinfo
  {journal} {Nanoscale Horiz.}\ }\textbf {\bibinfo {volume} {5}},\ \bibinfo
  {pages} {308} (\bibinfo {year} {2020})}\BibitemShut {NoStop}%
\bibitem [{\citenamefont {Sriv}\ \emph {et~al.}(2020)\citenamefont {Sriv},
  \citenamefont {Nguyen}, \citenamefont {Lee}, \citenamefont {Lim},
  \citenamefont {Nguyen}, \citenamefont {Kim}, \citenamefont {Cho},\ and\
  \citenamefont {Cheong}}]{Sriv2020}%
  \BibitemOpen
  \bibfield  {author} {\bibinfo {author} {\bibfnamefont {T.}~\bibnamefont
  {Sriv}}, \bibinfo {author} {\bibfnamefont {T.~M.~H.}\ \bibnamefont {Nguyen}},
  \bibinfo {author} {\bibfnamefont {Y.}~\bibnamefont {Lee}}, \bibinfo {author}
  {\bibfnamefont {S.~Y.}\ \bibnamefont {Lim}}, \bibinfo {author} {\bibfnamefont
  {V.~Q.}\ \bibnamefont {Nguyen}}, \bibinfo {author} {\bibfnamefont
  {K.}~\bibnamefont {Kim}}, \bibinfo {author} {\bibfnamefont {S.}~\bibnamefont
  {Cho}}, \ and\ \bibinfo {author} {\bibfnamefont {H.}~\bibnamefont {Cheong}},\
  }\href {\doibase 10.1038/s41598-020-68744-2} {\bibfield  {journal} {\bibinfo
  {journal} {Scientific Reports}\ }\textbf {\bibinfo {volume} {10}},\ \bibinfo
  {pages} {11761} (\bibinfo {year} {2020})}\BibitemShut {NoStop}%
\end{thebibliography}%

\end{document}